\documentclass[aps,pra,twocolumn,showpacs,superscriptaddress]{revtex4-1}
\usepackage{amsmath,amssymb}
\usepackage[matha,mathx]{mathabx}
\usepackage{txfonts}
\usepackage{float}
\usepackage[pdftex]{graphicx}
\usepackage{subfigure}
\usepackage{natbib}
\usepackage{bm}
\newcommand \beq{\begin{eqnarray}}
\newcommand \eeq{\end{eqnarray}}

\begin{document}

\title{Spinor Bose gas in an elongated trap}

\author{Shun Uchino}
\affiliation{DQMP, University of Geneva, 24 Quai Ernest-Ansermet,
1211 Geneva, Switzerland}

\date{\today}

\begin{abstract}
We examine a spinor Bose gas
confined by an elongated trap.
Since a spin-independent energy is much higher than
a spin-dependent energy in alkali species, the system exhibits
different properties by changing a radial confinement.
We show that if a spin-dependent coupling is positive,
a spin-liquid condensate,
which breaks the charge $U(1)$ symmetry but
preserves the spin rotational symmetry, can be realized
in an intermediate confinement regime. 
Properties of the spin-liquid condensate are visible 
if a temperature is lower than
a spin gap to characterize the spin-disorder property.
 If a temperature is higher than the gap
but lower than a spin-dependent coupling energy,
on the other hand,
a regime in which a spin sector is described by a semiclassical wave emerges.
A characterization in each regime by means of
correlation functions and topological solitons is also discussed.
\end{abstract}


\maketitle

\section{Introduction}
Quantum many-body physics in reduced dimension is fundamentally
different from that in higher dimensions \cite{giamarchi2003quantum}.
When it comes to bosonic systems,
whilst the presence of a Bose-Einstein condensate (BEC) 
\cite{pitaevskii2003bose} is
the natural consequence for the systems in higher dimensions,
there is no BEC in an interacting one dimensional system
due to quantum fluctuations
\cite{PhysRev.158.383,pitaevskii1991uncertainty}.
Instead in many cases
the so-called Tomonaga-Luttinger liquid (TLL) \cite{giamarchi2003quantum},
which is also responsible to describe many fermionic one dimensional systems,
is realized.

A well-known criterion for the emergence of 
quantum one dimensional systems is that 
temperature and interaction energy  should
 be much smaller than a confinement energy
 along higher dimensional directions \cite{giamarchi2003quantum}.
However, one may encounter a nontrivial low-dimensional system
if there are several energy scales each of which is energetically separated.
A typical example is a quasi-one dimensional superconductor discussed
in condensed matter physics where an energy scale of electrons
is much higher than that of Cooper pairs.
In this system, therefore, 
it is expected that by changing a radial confinement
properties of the superconductor change
despite the three dimensional motion in each electron
\cite{arutyunov2008superconductivity,bezryadin2012}.

Recently, on the other hand,
cold atoms have provided remarkable realizations 
of systems in reduced dimension \cite{cazalilla2004bosonizing,
RevModPhys.80.885,RevModPhys.83.1405}.
Practically speaking, such systems 
can be prepared by making a trap very anisotropic 
\cite{PhysRevLett.87.080403,PhysRevLett.87.130402,PhysRevLett.87.160405}
or by loading systems 
on two-dimensional optical lattices 
\cite{PhysRevLett.87.160405,PhysRevLett.91.250402,PhysRevLett.113.215301}.

In particular, one can consider 
a spinor Bose gas
realized in cold atoms
\cite{kawaguchi2012spinor,RevModPhys.85.1191}
as 
a system with 
multiple energy scales.
Experimentally,
this system   
has been realized with alkali species such
as $^{23}$Na and $^{87}$Rb where 
$s$-wave scattering lengths 
to characterize interatomic interactions take similar values
for different spin channels. This
implies that
a spin-dependent coupling is much smaller than a spin-independent 
coupling.
Therefore,
a variance of spin can easily be restricted 
to a one direction compared with that of charge,
which has indeed been confirmed in Ref. \cite{PhysRevLett.110.165301}.

\begin{figure}[t]
 \begin{center}
  \includegraphics[width=0.8\linewidth]{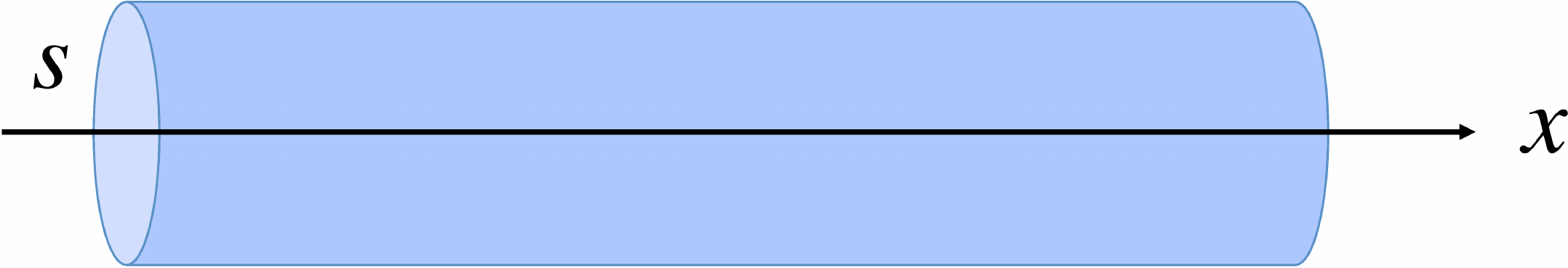}
  \caption{(color online) Spinor Bose gas 
in an elongated trap with a cross section $s$.}
  \label{fig1}
 \end{center}
\end{figure}

In this paper, we examine a
spinor Bose gas confined by
an elongated trap as illustrated in Fig. \ref{fig1} 
and show that the system experiences  
nontrivial properties by changing a radial confinement.
In particular, by considering a positive spin-dependent coupling,
we predict that in a region where
a charge sector is three dimensional while a spin sector
is one dimensional,
a spin-liquid condensate
\cite{PhysRevLett.88.163001,zhou2003spin,
PhysRevLett.97.120406,PhysRevA.84.053625,PhysRevLett.113.080402}
is realized.
We also discuss  temperature effects and show that a regime where
the spin sector is described by
a semiclassical wave 
\cite{PhysRevB.50.9265,PhysRevB.57.8307,PhysRevB.59.9285,sachdev2007quantum}
is achieved in certain temperature regime
although it is difficult to obtain such a regime in an
antiferromagnetic quantum spin-1  system \cite{PhysRevB.50.9265}.
This difference can be attributed to the fact that there are several length
scales in spinor bosons while there is only one length scale
in a quantum spin system.
A characterization in each phase by means of correlation functions
and vortices
is also discussed.

\section{Model}
We consider spin-1 bosons in an elongated trap.
The Hamiltonian with $U(1)\times SO(3)$ symmetry
is given by
\beq
H=\int d^3x \Big[\psi^{\dagger}_m\left(-\frac{\hbar^2\Delta}{2M}
+V(\mathbf{x})\right)\psi_m\nonumber\\
+\frac{c_0}{2}(\psi^{\dagger}_m\psi_m)^2
+\frac{c_1}{2}(\psi^{\dagger}_m\mathbf{F}_{mn}\psi_n)^2
\Big],
\eeq
where $\psi_m$ is the boson field with the mass $M$, $F_i$ $(i=x,y,z)$
are spin-1 matrices, and $V(\mathbf{x})$ is a confined potential
described by a harmonic oscillator.
In what follows, let us focus on a situation that
a confined potential is imposed only along the radial direction,
namely, $V(\mathbf{x})=\frac{M\omega^2_{\perp}}{2}(y^2+z^2)$.
In fact, such a trapping has already been realized 
in Ref. \cite{PhysRevA.71.041604}.
The spin-independent  and spin-dependent couplings 
are respectively expressed with $s$-wave scattering lengths
as
$c_0=4\pi\hbar^2(a_0+2a_2)/3M$ and $c_1=4\pi\hbar^2(a_2-a_0)/3M$,
where $a_f$ $(f=0,2)$ is
the s-wave scattering length of the total spin-$f$ channel.

\section{Spinor Bose gas in an elongated trap}
Here, we discuss a spin-1 Bose gas
for $c_1>0$ in different harmonic oscillator frequencies
at absolute
zero.

\subsection{3D regime}
We consider a regime obeying the following condition:
\beq
nc_0, nc_1 \gg \hbar\omega_{\perp}
\label{eq:3d}
\eeq
where $n$ is the density,
$nc_0$ and $nc_1$ denote the interaction entries with respect to
charge and spin, respectively. 
In this regime, since there are a number of transverse modes,
the system is treated as 3D and thereby
the Gross-Pitaevskii approximation is justified at least
for a weak-coupling case.
This approximation is nothing but the substitution of a $c$-number
field $\phi_m$ for the original field. 
Such a $c$-number field can be determined by the saddle point
equation known as the Gross-Pitaevskii equation.
It can be shown that for $c_1>0$ the ground state is a polar phase whose
order parameter is given by \cite{PhysRevLett.81.742,JPSJ.67.1822}
\beq
\langle\psi_m\rangle\equiv\phi_m=\sqrt{n}(0,1,0)^T,
\label{eq:op}
\eeq
where $T$ represents transposition.
We also note that in general $n$ has a dependence of $r_{\perp}$,
which may be approximated by the Thomas-Fermi profile due to 
the condition
\eqref{eq:3d} 
\cite{pitaevskii2003bose}.

Let us next look at a symmetry breaking.
From \eqref{eq:op}, we see that  $\langle
\mathbf{F}\rangle=\mathbf{0}$
and the charge $U(1)$ symmetry is spontaneously broken, which is
accompanied by an off-diagonal long-range order (ODLRO).
In addition, it is easy to show \cite{PhysRevA.81.063632}
\beq
\langle [\psi_{m'},\psi^{\dagger}_{m}(F_i)_{mn}\psi_n]\rangle\ne0,
\ \ \ (i=x,y)
\label{eq:ssb}
\eeq
which states that the spin rotational symmetry is also spontaneously broken
as $SO(3)\to SO(2)$.
An accurate order parameter manifold including discrete symmetry
can be obtained by operating $e^{i\theta}e^{iF_z\alpha}e^{iF_y\beta}
e^{iF_z\gamma}$ into Eq. \eqref{eq:op}, where $\theta$ is the phase of
the
charge $U(1)$
and $(\alpha,\beta,\gamma)$ are Euler angles, and
is shown to be \cite{PhysRevLett.87.080401}
\beq
G/H=[U(1)\times S^2]/Z_2.
\label{eq:opm}
\eeq
The above order parameter manifold physically represents that
while the charge $U(1)$ symmetry 
and rotational symmetry along the $x$ and $y$ axes
are broken, the
rotational symmetry along the $z$ and spin-charge coupled discrete 
$Z_2$ symmetry remain.

Due to the spontaneous symmetry breaking,
gapless modes known to be Nambu-Goldstone modes
emerge.
The number of Nambu-Goldstone modes is expected to 
be equal to the dimension of the order parameter manifold, which
is 3 in Eq. \eqref{eq:opm}.
In fact, we can check this forecast by means of the analysis
of collective modes.
To this end, we consider fluctuation effects from
the $c$-number field and
expand it
up to the linear order of the fluctuation field 
in the Gross-Pitaevskii equation,
which leads to the Bogoliubov equation.
By diagonalizing it,
we obtain the following Bogoliubov modes:
\cite{PhysRevLett.81.742,JPSJ.67.1822}
\beq
E_{c,\mathbf{k}}&=&\sqrt{\epsilon_{\mathbf{k}}(\epsilon_{\mathbf{k}}+2nc_0)},\\
E_{f_i,\mathbf{k}}&=&\sqrt{\epsilon_{\mathbf{k}}(\epsilon_{\mathbf{k}}+2nc_1)},
\ \ (i=x,y)
\eeq
where $\epsilon_{\mathbf{k}}=\hbar^2k^2/(2M)$,
 $E_{c,\mathbf{k}}$ denotes the Bogoliubov mode coming from 
charge fluctuations, and $E_{f_x,\mathbf{k}}$ and $E_{f_y,\mathbf{k}}$
are related to spin fluctuations.
Thus, it turns out that all of the above Bogoliubov modes are shown to be
linear gapless by taking the limit $k\to0$,
which is consistent with  the Nambu-Goldstone theorem.

\subsection{1D regime}
We next consider a situation that
atomic motions along the transverse directions
are frozen and the gas is kinematically 1D, which implies
\beq
nc_0,nc_1\ll \hbar\omega_{\perp}.
\label{eq:1d}
\eeq
In this case, the conventional 
Gross-Pitaevskii approach is no longer appropriate
due to the absence of a BEC.
Nevertheless, we can  employ a semiclassical analysis 
to describe low-energy properties in the system.

A key idea is an introduction of a number-phase representation in
bosons \cite{essler2009spin},
\beq
\psi_m=\sqrt{n}n_{m}e^{i\theta_m},
\label{eq:1dop}
\eeq
with a constraint $\sum_mn_m^2=1$.
If we discuss such a representation in spinless bosons,
there are two variables, which are density and phase. 
Since the density is a massive degree of freedom,
we can integrate out it as far as low-energy properties in the system
are concerned.
As a consequence,
we obtain a phase-only action, which is essentially the TLL 
whose spectrum is linear gapless.
This is indeed the correct low-energy effective action 
\cite{giamarchi2003quantum}.
For the case of spinor bosons, on the other hand
 while the situation
is much involved due to the multiple number-phase variables,
the technique used in spinless bosons is still applicable.
First, by solving saddle point equations in the number-phase representation,
we can specify massive and massless degrees of freedom at
the lowest order.
Second, by expanding the effective potential from the minimum
with respect to the massive degrees of freedom up to second order
and then integrating out them, we arrive at the following effective
action \cite{PhysRevLett.87.080401,essler2009spin}:
\beq
S&=&S_{TLL}+S_{NL\sigma M},\\
S_{TLL}&=&\frac{\hbar K_c}{2\pi}\int 
dtdx\Big[\frac{1}{v_c}(\partial_t\theta_+)^2
-v_c(\partial_x\theta_+)^2\Big],
\label{eq:tll}\\
S_{NL\sigma M}&=&\frac{\hbar}{2g}\int v_sdtdx\Big[(\partial_t\mathbf{m}/v_s)^2
-\{(\partial_x\mathbf{m})^2
\}
\Big].
\label{eq:nlsm}
\eeq
Here $S_{TLL}$ is the TLL action reflecting in
a superfluid property, which can be described by 
the velocity $v_c=\sqrt{\bar{n} \bar{c}_0/M}$ and the so-called
TLL parameter $K_c=\pi\hbar\sqrt{\bar{n}/(M\bar{c}_0)}$
with the line density $\bar{n}=ns$ and 
effective 1D coupling $\bar{c}_i=c_i/s$
\footnote{Here, we implicitly assume $a_{\perp}\gg a$,
which is well-satisfied in current experiments in the absence of 
a Feshbach resonance. If it is not the case, the 1D coupling should 
be renormalized based on \cite{PhysRevLett.81.938}}.
On the other hand, $S_{NL\sigma M}$ 
is the nonlinear $\sigma$ model action
describing the spin dynamics with velocity 
$v_s=\sqrt{\bar{n}\bar{c}_1/M}$,
 dimensionless coupling constant $g=\sqrt{M\bar{c}_1/(\hbar^2\bar{n})}$, 
and constraint $\mathbf{m}^2=1$.
The variables in the above action are associated with
\eqref{eq:1dop} via
$\theta_{\pm}=(\theta_1\pm\theta_{-1})/2$,
and
$\mathbf{m}=(\sin\theta\cos\theta_-,\sin\theta\sin\theta_-,\cos\theta)^T$
with
 $\theta=\sin^{-1}[(n_1-n_{-1})/\sqrt{2}]$.
Thus, we could obtain the action with the spin-charge separation,
which is naturally expected in one dimensional systems 
\cite{giamarchi2003quantum}.

Here, we look at excitation properties in each sector.
As for the charge sector, the gapless property of the spectrum
should be maintained unless there is a commensurate potential,
which introduces cosine terms and may make it gapful.
As for the spin sector described by the 1+1 dimensional 
nonlinear $\sigma$ model, the excitation is going to be
gapped. This is in contrast with the two or three dimensional cases
where the spin excitation is gapless due to the absence of infrared
divergences \cite{tsvelik2006quantum}. 
In fact, the above properties can be shown exactly for the $c_0=c_1$ case
\cite{cao2007paired}
where the so-called Takhatajan-Babujian limit 
\cite{takhtajan1982picture,babujian1982exact}
is realized and the Bethe ansatz solution is available. 
In addition, the corresponding analysis shows that 
the ground state consists of a string solution 
forming a spin-singlet pair, which is responsible for
the spin gap.

We now come back to the semiclassical analysis to see correlation
properties.
The above
semiclassical analysis indicates that
the bosonic field can be expressed as
\beq
\psi_m\approx \sqrt{n}e^{i\theta_+}((m_1+im_2)/\sqrt{2},
m_3,-(m_1-im_2)/\sqrt{2})^T.\nonumber\\
\label{eq:semiclassical-psi}
\eeq
Thus, we see that 
the one-particle correlation function in a long range
decays exponentially
since
$\langle m_i(r)m_j(0)\rangle\sim\delta_{i,j}e^{-\Delta_sr/(\hbar
v_s)}$, where $\Delta_s$ is the spin gap originating from
the nonlinear $\sigma$ model action \eqref{eq:nlsm}, 
while a correlator of the charge sector
decays algebraically as $\langle
e^{i\theta_{+}(r)}e^{-i\theta_+(0)}\rangle
\sim r^{-1/(2K_c)}$,
 where $r=\sqrt{x^2+v_j^2t^2}$ ($j=c$ or $s$).
Thereby, 
the dominant correlation turns out to be the following 
spin-singlet pair correlation:
\beq
\langle P_0^{\dagger}(r) P_0(0)
\rangle\approx \langle e^{-2i\theta_+(r)}e^{2i\theta_+(0)}
\rangle\sim r^{-2/K_c},
\label{eq:tl-correlation}
\eeq 
where
$P_0=2\psi_1\psi_{-1}-\psi_0\psi_0$.
As a consequence, it follows that
 the ground-state of a one-dimensional 
spinor Bose gas for $c_1>0$
is the spin-liquid TLL.

\subsection{Intermediate regime}

In accord with the analyses discussed above, 
we next consider an intermediate confinement regime defined by
\beq
nc_1\ll\hbar\omega_{\perp}\ll nc_0.
\label{eq:intermediate-r}
\eeq
The above states that the charge sector is three dimensional while
the spin sector is one dimensional, that is, in the latter sector 
a quantum fluctuation is important.

To consider this regime, we come back to
the number-phase representation of bosons
discussed in the one dimensional regime.
An important observation is that the number-phase representation of
Eq. \eqref{eq:semiclassical-psi} is still applicable
for higher dimensional cases.
Then, the crucial point is that for higher dimensions
a quantum fluctuation effect trying to break the order
is weak enough, which ensures that 
excitations both from the charge and spin are gapless, and
leads to the ODLRO in the system.
Thus, for higher dimensions Eq. \eqref{eq:semiclassical-psi} 
corresponds to \eqref{eq:op} up to $U(1)\times SO(3)$ rotations.

On the other hand, from the condition
\eqref{eq:intermediate-r},
the spin correlation length $\xi_s=\hbar/\sqrt{2Mnc_1}$
providing a typical length scale in the spin sector
is larger than the radial oscillator length $a_{\perp}=
\sqrt{\hbar/M\omega_{\perp}}$ while the charge correlation length
$\xi_c=\hbar/\sqrt{2Mnc_0}$ as a typical length scale in the charge sector
is smaller than $a_{\perp}$.
This implies
that the effective low-energy action describing the intermediate regime
is given by
\beq
&&S=S_c+S_{NL\sigma M},\\
&&S_c=\hbar^2\int
dtd^3x\sqrt{\frac{n}{4Mc_0}}
\Big[\frac{1}{v_c}(\partial_t\theta_+)^2
-v_c(\nabla\theta_+)^2
\Big],
\eeq
where $S_{NL\sigma M}$ corresponds to Eq. \eqref{eq:nlsm}.
Since the effective action of the charge sector is three dimensional,
a quantum fluctuation effect is significantly reduced
to cause the emergence of a BEC.
As in the case of the 3D regime, the density profile may be captured by the 
Thomas-Fermi
one due to $nc_0\gg\hbar\omega_{\perp}$.
On the other hand, the effective action of the spin sector 
is the 1+1 dimensional nonlinear $\sigma$ model and therefore,
the spin-singlet formation and corresponding spin gap
emerge.

Now, we look at correlation properties in this regime.
A striking property is that the one-particle correlation function
decays exponentially as in the case of the 1D regime, 
which implies the absence of an ODLRO of the one-particle density matrix.
This is due to the fact that while
$\langle e^{-i\theta_+(r)}e^{i\theta_+(0)}\rangle$ is a constant
even in a long range, the one-particle correlation itself
disappears in the corresponding limit since 
$\langle m_i(r)m_j(0)\rangle\sim \delta_{i,j}e^{-\Delta_sr/(\hbar v_s)}$.
At the same time, this property does not mean the absence of an
ODLRO in the system itself.
In fact, the spin-singlet pair correlation function,
$\langle P_0^{\dagger}(r)P_0(0) \rangle$ 
remains nonzero in a long range.
Namely, as in the case of fermionic superfluids described
by the BCS theory, the ODLRO comes out from
the two-particle density matrix.

Here we point out that the BEC realized in this intermediate regime
can be regarded as a spin-liquid condensate discussed
in Refs. \cite{PhysRevLett.88.163001,zhou2003spin,
PhysRevLett.97.120406,PhysRevA.84.053625,PhysRevLett.113.080402}.
This is a bosonic state such that 
the spin rotational symmetry remains unbroken
while the $U(1)$ charge symmetry breaks spontaneously.
In our model, the $U(1)$ symmetry breaking causes
the ODLRO of the two-particle density matrix
not of the one-particle density matrix due to the spin-disorder property.
We notice the difference between the spin-liquid condensate and
the spin-singlet pair condensate obtained  
with a single-mode approximation 
\cite{PhysRevLett.81.5257,PhysRevLett.84.1066,van2007bose,PhysRevA.78.023609}.
In the latter case, 
while the spin rotational symmetry is maintained,
 the ODLRO of the one-particle density
matrix exists since such a singlet-pair formation occurs between
bosons with zero momentum.
An emergence of 
bosonic condensates without the ODLRO
of the one-particle density matrix is considered to be unusual.

\section{Discussion}
\subsection{Characterization in each phase}
Now, we wish to discuss a characterization in each phase
in light of experimental observables.
To this end, we focus on correlation properties and 
vortices, both of which can be measured in experiments.
When it comes to correlation properties,
the one-particle density matrix to confirm
an ODLRO in a spinless BEC
has been measured in \cite{bloch2000measurement,PhysRevLett.98.090402},
and pair correlation function to confirm existence of 
pair condensation has been measured in \cite{PhysRevLett.94.110401}
via the atom shot noise in absorption imaging.
On the other hand, the several topological excitations including vortices 
in a spinor BEC have been reported in Refs.
\cite{PhysRevLett.89.190403,PhysRevLett.90.140403,PhysRevA.73.063605,
PhysRevA.77.041601,PhysRevLett.102.030405,sadler2006spontaneous,
PhysRevLett.108.035301,PhysRevLett.111.245301}.

In the three dimensional regime \eqref{eq:3d}, whose existence 
has been confirmed by the experiment \cite{stenger1998spin}, 
the ODLRO of the one-particle density matrix
in the $m=0$ component should appear
in the order parameter \eqref{eq:op}, 
In addition, since the system size is bigger than the charge and spin
correlation lengths, both of integer and half-quantized vortices 
are topologically allowed due to $H=Z_2\ltimes U(1)$ and $\pi_{1}(G/H)=Z$
in the polar phase, where $\ltimes$ and $\pi_1$ mean
the semidirect product and first homotopy group, respectively. 

In the one dimensional regime \eqref{eq:1d},
vortex excitations are forbidden due to $a_{\perp}< \xi_c,
\xi_s$. 
The one-particle density matrix decays exponentially 
in contrast with the three dimensional case.
At the same time, since the charge part can be described by
the TLL, we expect a quasi-long range order
of the pair correlation function as Eq. 
\eqref{eq:tl-correlation}.

In the intermediate regime \eqref{eq:intermediate-r},
an integer quantum vortex is allowed although a half-quantum 
vortex is forbidden.
This is due to the fact that there is (no) the rotational degree of freedom 
along the radial direction in the charge (spin) sector since
$\xi_c< a_{\perp}< \xi_s$.
The one-particle correlation decays exponentially as in the case of
the one dimensional regime.
On the other hand, the pair correlation function acquires the 
ODLRO because of the three dimensional properties in the charge sector.

\subsection{Temperature effect}
We turn to consider effects of a finite temperature
within a range where the effective theory discussed above is applicable.
This implies that the thermal lengths
($\hbar v_c/T$ and $\hbar v_s/T$ )
should be longer than the 
cut-off length of the theory in each sector, which  
is of the order of the healing lengths 
($\xi_c$ and $\xi_s$).
In other words, we wish to consider the temperature satisfying
$T< nc_0, nc_1$.

Although in the 3D regime we do not expect any modification
of the argument at least in this temperature regime,
we need to care about temperature effects in the one dimensional 
and intermediate regimes.
This is because in both cases there is the spin gap $\Delta_s$ as 
an additional energy scale in the spin sector.
The typical
behavior of the spin gap is known to be $\Delta_s\sim
nc_1 e^{-2\pi/g}$
in the weak-coupling limit, and
is $\Delta_s\sim nc_1$ in the strong-coupling limit \cite{shlyapnikov2011polar}.
The dimensionless coupling $g$ 
is expected to be small in typical experiments of cold atoms 
and therefore $\Delta_s$ can be much smaller than
$nc_1$, which is in contrast with an antiferromagnetic
quantum spin-1 system where a spin gap is of the order of an 
exchange energy \cite{sachdev2007quantum}.
This is due to the fact that although an exchange energy is essentially
only one energy scale in such a spin system,
$g$ in a spinor gas can be determined by the competition
among the three different length scales: $a_{2}-a_0$,
$a_{\perp}$, $1/n^{1/3}$. 
On the other hand, 
the realization of the strong-coupling regime would be important to
obtain $\Delta_s>T$, which ensures the spin-liquid properties discussed
above. To this end, optical lattice technique,
optical Feshbach resonance \cite{PhysRevA.79.023401},
or microwave Feshbach resonance \cite{PhysRevA.81.041603} 
to tune the spin-dependent
interaction may be promising.

At the same time, even when the weak-coupling
regime is realized and $\Delta_s< T$ is met,
we have a chance to obtain an 
interesting regime $\Delta_s< T <nc_0,nc_1$.
Due to $ T< nc_1$, the usage of the nonlinear $\sigma$ model
to describe the low-energy property of the spin sector is still
reasonable.
Then, it turns out that the spin sector is  described by a semiclassical wave,
which has been originally
discussed in quantum spin systems
\cite{PhysRevB.50.9265,PhysRevB.57.8307,PhysRevB.59.9285,sachdev2007quantum}.
In this regime, a decay of a two-point correlation function in spin
$\langle m_i(r) m_j(0)\rangle$
can be
characterized by a correlation length 
$\xi\sim \frac{\hbar v_s}{T}\log(T/\Delta_s)$.
An important point is that a crossover with respect to
 different behaviours of the correlator occurs around $\xi$, that is,
the decay can be described by algebraic one for $|r|\lesssim\xi$
while it can be described by exponential one for $r\gg \xi$
\cite{giamarchi2003quantum}.
Thus, the properties of this regime can be again captured by means of
the correlation function.
 
\subsection{Species} 
So far, spin-1 bosons for $c_1>0$ have been discussed, which have
been realized 
in $^{23}$Na   \cite{stenger1998spin}.
In this case, the ratio of the spin-dependent coupling to the
spin-independent coupling $c_1/c_0$ is indeed small and is of the 
order of $10^{-2}$, which may open a window of opportunity to see 
the properties discussed above.

We can also consider $^{87}$Rb atoms where
the ferromagnetic condensate has been realized in the three
dimensional regime since $c_1<0$ \cite{PhysRevLett.92.140403}.
In this case, the ratio $|c_1|/c_0$ is of the order of $10^{-3}$,
which is smaller than the spin-1 $^{23}$Na case.
At the same time, the change of properties for $c_1<0$
may not be drastic compared with that for $c_1>0$
since it is known that a ferromagnetic property always
maintains 
even in one dimensional case \cite{PhysRevLett.110.130405}.

In addition,
a spin-2 BEC has also been realized with $^{87}$Rb
\cite{PhysRevLett.92.140403,PhysRevLett.92.040402}.
In the spin-2 case, we need an additional case since
there are two-independent spin-dependent couplings,
$c_1$ and $c_2$ each of which describes the spin-spin coupling and
pair-singlet coupling, respectively.
For a spin-2 $^{87}$Rb BEC realized in 3D,
the ground state is expected to be a nematic phase,
which is realized for $c_1>0$ and $c_2<0$.
The ratios $c_1/c_0$ and $|c_2|/c_0$ are respectively 
of the order of $10^{-2}$ and $10^{-3}$.
The nematic phase has similar properties as the polar phase
in the sense that there is no magnetization and the spin-singlet
amplitude takes a nonzero value in the ground state
\cite{PhysRevLett.84.1066,PhysRevA.61.033607,PhysRevA.65.063602}.
However, the nematic phase has an unusual property since
it has an accidental $SO(5)$ symmetry at the semiclassical level, 
which leads to the emergence of  quasi-Nambu-Goldstone
modes \cite{PhysRevLett.105.230406}.
Thus, such modes gain  masses due to the explicit symmetry breaking
from the interaction term with $SO(3)$
via the quantum fluctuations even in 3D. 
When it comes to the reduced dimensional case, however,
the spin-singlet projection operator possessing
 $SO(5)$ symmetry in the spin-2 case are responsible for
spin gaps and then, $SO(3)$ symmetric interaction may just cause
the renormalization on the gaps \cite{shlyapnikov2011polar}.
Thus, as in the case of the polar phase in a spin-1 BEC,
the excitation spectra except for the charge sector obtain  gaps
in such a spin-2 Bose gas in 1D. 

In addition, the ratios 
of a spin-dependent coupling to a spin-independent coupling
may be changed
by using the optical \cite{PhysRevA.79.023401}
and microwave Feshbach resonances \cite{PhysRevA.81.041603}. 

\section{Summary}
We have discussed spinor bosons confined by an elongated trap
and shown that the system experiences 
different properties by changing a radial confinement.
At absolute zero, we have predicted the spin liquid condensate phase realized 
in an intermediate regime where the charge sector is 3D
while the spin sector is 1D.
In this condensate, the charge $U(1)$ symmetry is spontaneously broken 
but the spin rotational symmetry is unbroken due to a spin gap.
We have pointed out that at 
a finite temperature higher than the spin gap but lower than
a spin-dependent coupling,
there is the regime where
the spin sector can be described by a semiclassical wave.
We suggest that each phase can be characterized by properties
of correlation functions and topological defects.

We also comment on a possible experiment to probe the phases discussed
above. The simplest way would be to measure the spin gap since
each phase can be discussed by existence or non-existene of the gap
or its magnitude as shown in the previous sections.
Then, 
magnon contrast interferometer recently 
demonstrated to measure a magnon gap in Ref. \cite{PhysRevLett.113.155302},
magnetic resonance spectroscopy  \cite{PhysRevA.87.061604},
or Bragg spectroscopy in a spin-selective manner
\cite{pitaevskii2003bose}
may be utilized to measure the spin gap.

It would also be interesting to extend our study to
higher spin cases.

\section*{acknowledgement}
The author  acknowledges
 T. Giamarchi for many stimulating discussions.
This work was supported by the Swiss National Science Foundation under
division II.

%

\end{document}